\documentclass[11pt]{article}

\usepackage[margin=0.78in]{geometry}
\usepackage[T1]{fontenc}
\usepackage{lmodern}
\usepackage{microtype}
\usepackage{amsmath,amssymb,amsthm,mathtools}
\usepackage{booktabs}
\usepackage{setspace}
\usepackage{natbib}
\usepackage{multibib}
\usepackage[hidelinks]{hyperref}
\setcitestyle{authoryear,round,aysep={,}}
\newcites{rep}{Replicated Papers}
\usepackage{graphicx}
\usepackage{caption}
\usepackage{etoolbox}
\newlength{\exhibitnotesskip}
\AtBeginEnvironment{figure}{\setlength{\exhibitnotesskip}{4pt}}
\newcommand{\exhibitnotes}[1]{%
  \par\vspace{\exhibitnotesskip}%
  \begin{minipage}{\linewidth}%
  \footnotesize\singlespacing\setlength{\parindent}{0pt}%
  \textit{Notes:} #1%
  \end{minipage}%
}

\newcommand{\E}{\mathbb E}

\newgeometry{margin=1.25in}

\title{How Different are Different Differences-in-Differences?\\ A Large-Scale Meta-Analysis}
\author{Peter Hull\thanks{Brown University Department of Economics; peter\_hull@brown.edu.\\ I thank Jason Abaluck, Zach Bleemer, Kirill Borusyak, Jiafeng Chen, Xavier Jaravel, Lihua Lei, Jonathan Roth, Pedro Sant'Anna, Jann Spiess, and Amanda Starc for helpful comments; Samuel Bazzi, Judson Boomhower, Enrico Cantoni, Pedro Dal Bó, Kyle Emerick, Luiz Felipe Fontes, Sebastian Galiani, Álvaro García-Marín, Matthew Gibson, Ricard Gil, Alex Hollingsworth, Aljoscha Janssen, Dean Karlan, Victor Lavy, Monica Martinez-Bravo, Benjamin Marx, Mathilde Muñoz, Peter Nencka, Mounu Prem, Paul Raschky, Jonah Rexer, Edson Severnini, Joakim Weill, and Thomas Wollmann for help replicating their papers; and OpenAI's ChatGPT for excellent research assistance. Refine.ink was used to check for consistency and clarity. }}
\date{November 2026}

\begin{document}
\maketitle

\begin{abstract}
\noindent I use AI-assisted replication to assess the empirical importance of negative weights in two-way fixed effects (TWFE) estimation, analyzing 1,430 staggered-adoption specifications from 116 economics papers. TWFE estimates are highly correlated with modern difference-in-differences estimates, with sign reversals in 12.0–14.9\% of comparisons and greater precision for TWFE. Hierarchical models accounting for estimation noise imply estimand correlations of 91–97\% and a 0.7–2.4\% probability of a statistically significant TWFE estimate opposing the modern estimand’s sign. These findings suggest negative weights may be a less serious problem in practice and illustrate how AI can support systematic, large-scale empirical assessments of new econometric methods.
\end{abstract}

\newpage
\onehalfspacing

\section{Introduction}

Difference-in-differences (DiD)---one of the most widely used empirical strategies in applied microeconomics \citep{goldsmithpinkham2026}---has recently undergone a crisis of faith. In staggered adoption settings, where a binary treatment is received at different calendar dates, the workhorse two-way fixed effects (TWFE) regression estimator is now understood to carry a ``negative weights'' concern. The concern is TWFE can apply non-convex weights to heterogeneous treatment effects when aggregating over event times, potentially leading to sign reversals. Even when all treatment effects are positive, the TWFE estimand can wind up negative---bringing into question the general reliability of the estimator.

This important theoretical development notwithstanding, the actual prevalence and impact of negative weights are ultimately empirical questions.  In many staggered adoption settings, TWFE weights may in fact be convex or nearly so. And even given a large share of negative weights, TWFE bias may be minimal when treatment effect variation is small or idiosyncratic. Switching from TWFE to a modern DiD estimator, which targets an unweighted average treatment effect, may then sacrifice the relative precision of regression estimates for robustness to a minor threat of sign reversals or other biases.

This paper studies the empirical importance of negative weights in TWFE estimation, using a large sample of replicated specifications from ten general-interest journals in economics. I first use agentic artificial intelligence (AI) to search 16,094 papers in 2000-2026 for eligible staggered-adoption DiD specifications with public replication archives. This search results in 1,430 replicated specifications from 116 papers, to which I then apply TWFE and a modern estimator. I consider both DiD-averaging (as in \cite{callawaysantanna2021}), which combines clean event-time DiDs with prespecified convex weights, and imputation (as in \cite{borusyakjaravelspiess2024}), which constructs counterfactual outcomes for individual treated units. Comparisons between conventional TWFE estimates and these modern DiD estimators give an indirect test of how much  negative weights affect the empirical conclusions.\footnote{Estimates may also flip signs because of estimation error or estimand differences due to, e.g., failures of the key parallel trends assumption. The analysis here accounts for the former but not the latter.}

Overall, the findings suggest negative weights may be a less serious problem in practice than in theory.  TWFE and modern DiD estimates are highly correlated, around 90\%, and sign reversals are fairly uncommon. Around 14.9\% of TWFE estimates change sign when switching to DiD-averaging; for imputation, this share is 12.0\%. Some of this represents noise: shares of sign flips with a statistically significant difference in estimates are a lower 3.9\% for DiD-averaging and 2.8\% for imputation. DiD-averaging and imputation yield standard errors that are, on average, 40\% and 9\% larger than TWFE's. I also find that negative weights are relatively infrequent and tend to make up a small share of the total absolute weight that TWFE applies to treatment effects.

There is, nevertheless, predictable variation in sign reversals. Later publication years (especially 2021 and after) and a paper's reporting of both TWFE and modern DiD estimators are associated with lower rates of sign reversals. Consistent with econometric theory, specifications with larger always-treated shares exhibit significantly more sign reversals for DiD-averaging while larger never-treated shares predict significantly fewer reversals for imputation. In contrast, journal rank, paper citation counts, and the prominence of specifications show no statistically significant association with sign reversals.

Finally, I use a hierarchical model to estimate the distribution of TWFE and modern DiD estimands while accounting for typically nontrivial estimation error. Model estimates confirm high ($91-97\%$) correlation across estimands and low (6.0-11.8\%) prior probabilities of sign reversals. I further use the model to estimate the probability of a ``significant'' sign reversal: the probability that a true positive (or negative) modern DiD estimand produces a statistically significantly negative (or positive) TWFE estimate. These rates are 2.4\% for DiD-averaging and 0.7\% for imputation. I obtain similar findings in the baseline Gaussian model and a nonparametric model, both allowing for precision dependence \citep{chen2026close}.

This paper builds on a recent and impactful econometrics literature studying negative weights and related issues in popular regression estimators. Examples include \cite{goodmanbacon2021}, \cite{dechaisemartindhaultfoeuille2020}, \cite{sunabraham2021}, \cite{callawaysantanna2021}, \cite{borusyakjaravelspiess2024}, \cite{goldsmithpinkhametal2024}, \cite{borusyakhull2024}, \citet{arkhangelskyetal2024designrobust},  \cite{wooldridge2025mundlak}, and \cite{gardneretal2026};  see  \citet{rothetal2023} and \citet{bakeretal2026} for recent reviews. I complement this work with a systematic empirical assessment of the negative weight concern across a large sample of economics papers.\footnote{Other potentially important issues raised in this literature---such as with time-varying controls, continuous treatments, extrapolation or underidentification in event study models, and choices of binning or other treatment specifications---are outside the scope of the present analysis. I deliberately focus on simple TWFE specifications with a binary treatment and no time varying controls to discipline the analysis. }

An important precedent for this paper is \citet{chiuetal2026}, who reanalyze 49 papers published in three leading political science journals during 2017–2023 using heterogeneity-robust estimators, parallel-trends diagnostics, and sensitivity analyses. My analysis covers more papers, journals, and specifications while focusing on staggered adoption settings (which characterize only 13 studies in \citet{chiuetal2026}). Beyond comparing estimates, I directly measure the prevalence and mass of negative TWFE weights and examine how sign reversals vary with publication timing, treatment composition, and specification characteristics. The estimated model allows us to distinguish between latent estimand reversals and reversals from estimation noise, and to quantify the empirically relevant probability of obtaining a statistically significant wrong-signed TWFE estimate.

Methodologically, this paper shows how modern AI tools can be used to conduct large-scale meta-analyses and explore the empirical content of econometric theory. While recent meta-analyses in economics---such as \citet{meager2019microcredit} or \citet{andrewskasy2019publication}---demonstrate the value of synthesizing evidence across studies, many such analyses use relatively small  samples of papers that are painstakingly collected and harmonized. I assembled a replication sample of around 100 papers in roughly two weeks on a personal laptop. Human prompting, judgment, and oversight remained essential, but agents performed the bulk of work in reading articles, identifying qualifying specifications, and running and debugging replication code. Similar protocols could make systematic empirical assessments of new econometric methods a more routine complement to theoretical advances, with further advances in AI technology expanding the scale and range of such assessments.\footnote{See, e.g., \citet{chen2026potentialweights} who similarly uses AI agents to survey empirical methods in 1,051 economics papers and separately reanalyzes nine applications. }

\section{Sample Construction}\label{sec:sample}
\begin{figure}[t!]
\centering
\caption{Papers in the Replication Sample}
\label{fig:sample-composition}
\includegraphics[width=\linewidth,trim=0pt 18pt 0pt 12pt,clip]{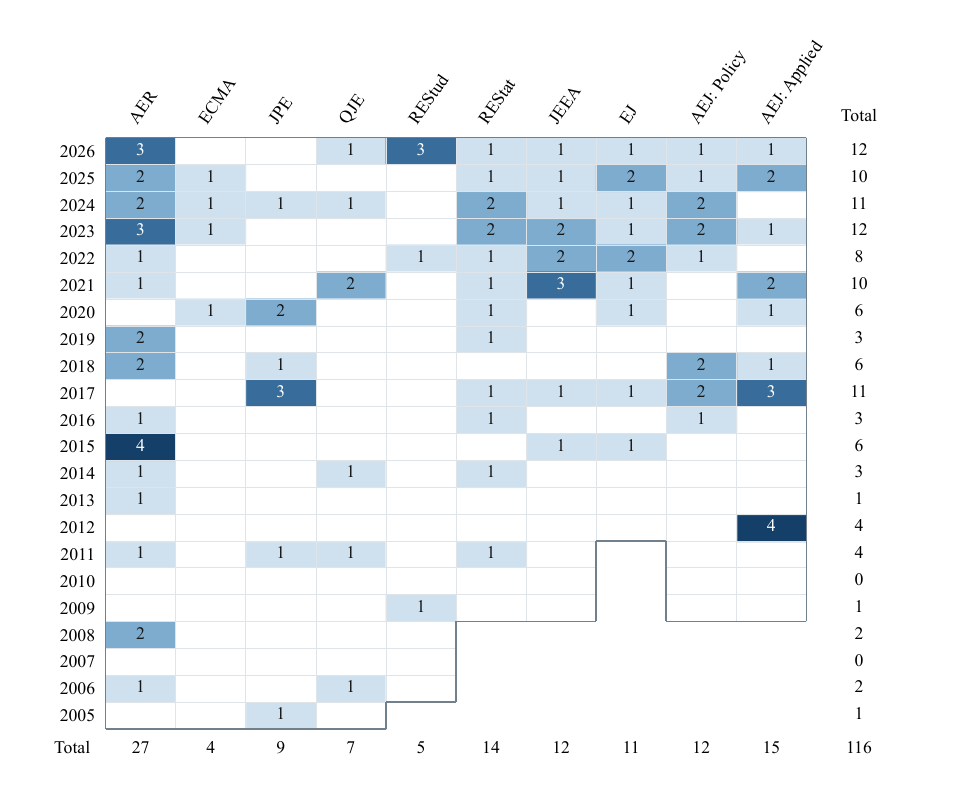}
\exhibitnotes{This figure counts papers in the replication sample. The final year includes papers that were published or forthcoming by October 10, 2026. White areas without cell outlines precede the journal search window.}
\end{figure}

Using agentic AI, I searched 16,094 bibliographic records from ten general-interest economics journals since 2000; this count includes nonresearch materials dropped during screening.\footnote{I used the GPT-6.1 Sol and GPT-6 Astra models of OpenAI's Codex. I am grateful to Drew Johnston of OpenAI for a free subscription to ChatGPT Pro during the month of October 2026.}  The journals are: the American Economic Review (AER), Econometrica (ECMA), the Journal of Political Economy (JPE), the Quarterly Journal of Economics (QJE), the Review of Economic Studies (REStud), the Review of Economics and Statistics (REStat), the Journal of the European Economic Association (JEEA), the Economic Journal (EJ), the American Economic Journal: Economic Policy (AEJ: Policy), and the American Economic Journal: Applied Economics (AEJ: Applied). Years were dropped prior to a journal's stated start of a public replication archive; Figure  \ref{fig:sample-composition} illustrates journal-specific sampling frames.\footnote{Papers had to be published or forthcoming by October 10, 2026 to be included in the search.} 

An initial screening dropped nonresearch materials such as journal frontmatter. We (I and the AI agents) used abstracts, keywords, and available article text and replication materials to identify potentially qualifying staggered-adoption specifications. Candidates then received further review of main-text tables, figures, equations, and numerical prose, with qualifying results linked to the corresponding code and data. The agents downloaded any publicly available copies of the article, keeping published versions where available; I manually downloaded the published version of any article or replication code/data requested by the agents for further screening, using valid credentials. 

Papers were retained if they satisfied two requirements. First, the paper must report an estimate from an eligible DiD specification in the main text, including tables, figures, and text discussions. Eligible specifications include TWFE, the \cite{callawaysantanna2021} DiD-averaging estimator, or the \cite{borusyakjaravelspiess2024} imputation estimator, used either statically or in an event study, with staggered adoption (i.e. multiple treatment adoption periods).\footnote{For a small share of specifications involving a reversing treatment, I omit post-adoption observations in which the original treatment indicator equals zero when constructing TWFE and modern DiD comparisons.} To ease comparison across methods, I excluded specifications with time-varying controls, unit trends, or other complications (see Appendix \ref{app:screening}). The second requirement is replicability of an eligible specification. Our replication primarily used public code/data; in some cases, authors provided additional materials from private correspondence. In the vast majority of cases replication was exact; deviations in other specifications are generally small  (see again Appendix \ref{app:screening}). We included all eligible and replicable specifications in a paper.

The search concluded with 1,430 specifications across 116 papers, with each journal and almost all years after 2004 represented. Figure  \ref{fig:sample-composition} shows journal-year paper counts and Appendix Table \ref{tab:paper-inventory} lists the papers themselves. Column 1 of Table \ref{tab:descriptive-statistics} summarizes this sample. Around half of all specifications come from a paper published in the so-called ``Top Five'' journals (AER, ECMA, JPE, QJE, and REStud), with 53\% published in 2021 or later. The average number of citations to a retained paper is 146 (specification-weighted). Around 27\% of specifications are event studies; around 23\% come from papers that report both TWFE and either a DiD-averaging or imputation estimator.\footnote{I do not find evidence for publication bias in the replication sample: e.g., estimated jumps in the density of
t-statistics are statistically insignificant at both the specification level (p=0.52) and the paper level (p=0.35).} 

\begin{table}[t!]
\centering
\caption{Replication Sample Summary}
\label{tab:descriptive-statistics}
\small
\setlength{\tabcolsep}{6pt}
\begin{tabular}{@{}lccc@{}}
\toprule
 & All & DiD-Averaging & Imputation \\
 & specifications & comparisons & comparisons \\
 & (1) & (2) & (3) \\
\midrule
Publication year & 2020.2 & 2020.6 & 2020.2 \\
Published in 2021 or later & 0.532 & 0.582 & 0.547 \\
Reports both TWFE and CSA/BJS & 0.234 & 0.241 & 0.237 \\
Journal rank & 5.637 & 5.758 & 5.585 \\
Top-five journal & 0.475 & 0.436 & 0.469 \\
Paper citations & 145.9 & 147.9 & 156.7 \\
Abstract/intro. prominence & 0.699 & 0.737 & 0.739 \\
Preferred specification & 0.569 & 0.598 & 0.583 \\
Event study & 0.274 & 0.306 & 0.311 \\
Always-treated share & 0.071 & 0.069 & 0.071 \\
Never-treated share & 0.402 & 0.384 & 0.394 \\
\midrule
Specifications & 1,430 & 997 & 1,060 \\
Comparisons & -- & 1,205 & 1,254 \\
Papers & 116 & 90 & 97 \\
\bottomrule
\end{tabular}
\exhibitnotes{This table summarizes specifications in the replication sample. Column 1 covers all specifications, while columns 2 and 3 cover specifications contributing at least one retained DiD-averaging or imputation comparison, respectively. An event study is counted once even when it contributes both pre- and post-treatment comparisons. Entries other than publication year, journal rank, and citations are proportions. Journal ranks are 1--10 in RePEc order, with 1 denoting the highest standing. The top-five journals are AER, ECMA, QJE, JPE, and REStud. Citations are published-paper OpenAlex counts retrieved in October 2026. Reports-both indicates the paper reported both TWFE and either a DiD-averaged or imputation estimator in the main text or appendix. Abstract/introduction prominence indicates that a result is mentioned in either section. In columns 1--3, the preferred specification indicator is nonmissing for 1,134, 788, and 837 specifications; the never-treated measure is nonmissing for 1,424, 997, and 1,060 specifications.}
\end{table}

Four other measures were constructed to assess specifications and the potential for sign reversals. The first is an indicator for whether a finding is mentioned in the paper abstract or introduction; Table \ref{tab:descriptive-statistics} shows 70\% have such prominence. The second is an indicator for whether the specification is ``preferred'': either explicitly described in the paper as a baseline, main, or benchmark specification or implicitly judged as one by the AI agents' full-text review. Among the 1,134 specifications able to be classified in this way, the table shows 57\% are preferred. Finally, we construct measures of the share of units in the specification which are always-treated or never-treated. As reviewed below, these shares are expected to drive differences between TWFE and the modern DiD estimators.

Columns 2 and 3 of Table \ref{tab:descriptive-statistics} summarize these variables in the subsamples for which direct comparisons between TWFE and one of the two modern DiD estimators can be made. These exclude specifications that yield extremely high-leverage estimates: those with fewer than 10 effective clusters or a maximum cluster share of more than 25\%.  Such specifications are likely to suffer from unstable estimates and unreliable inference, complicating the analysis. Reassuringly, all specification characteristics are similar in both subsets, which represent large majorities of papers (77\% and 83\%, respectively). For event studies, we average post-event and pre-event estimates to obtain up to two comparisons per specification.

In total, we obtain 1,205 comparisons between TWFE and DiD-averaging and 1,254 comparisons between TWFE and imputation; see Appendix Table \ref{tab:paper-inventory} for paper-specific counts. For comparability across specifications, I normalize all outcomes to have unit standard deviation. I next review the estimators and the leading negative-weight concern in TWFE which the modern DiD estimators are designed to address.

\section{TWFE and Modern DiD}\label{sec:methods}

f
For each specification, the TWFE estimate  $\hat\beta^{TWFE}$ comes from a regression of the form:
\begin{align}
Y_{j} = \alpha_{u(j)} + \lambda_{t(j)} + D_j^\prime \beta^{TWFE} + W_{u(j)}^\prime\gamma_{t(j)} + \varepsilon_{j}.\label{eq:twfe}
\end{align}
Here $Y_j$ is the outcome for observation $j$, $u(j)$ indexes a stable unit or group, $t(j)$ indexes calendar time, $D_j$ is a treatment vector discussed below and $W_{u(j)}$ contains fixed unit characteristics or subgroup indicators (and may be empty). The $u(j)$ and $t(j)$ notation accommodates both panels and repeated cross-sections; for a panel indexed by unit $i$ and time $t$, we have $j=(i,t)$, $u(j)=i$, and $t(j)=t$.\footnote{Throughout I include original analysis weights; I abstract away from them here to keep notation simple. Seasonal calendar indicators are also permitted (see Appendix \ref{app:screening}) and are again suppressed to ease notation.}

In static TWFE specifications, $D_j$ is a single indicator that switches on at treatment adoption. For example, \citetrep{martinezbravoetal2022} study the effect of introducing village elections in China on outcomes such as public-goods expenditure and exemptions from the one-child policy. Here $W_u$ contains province indicators, allowing for province-specific year effects. In event-study specifications, $D_j$ is a vector of (possibly binned) relative event-time indicators with a reference period omitted. For example, \citetrep{baileygoodmanbacon2015} estimate the dynamic effects of establishing community health centers on age-adjusted mortality, using post-establishment coefficients to estimate treatment effects and pre-establishment coefficients to assess pretreatment trends. As noted above, I separately average the available pre- and post-treatment coefficient estimates over common supported windows for the comparisons with modern estimators.

A recent concern with TWFE is the potential for sign reversals: the regression estimand can be negative even when every individual treatment effect is positive, because it can attach negative weights to some treated observations \citep{goodmanbacon2021,dechaisemartindhaultfoeuille2020}. Formally, under the usual no-anticipation and parallel-trends assumptions (here given $W_{u(j)}$), the static TWFE estimand can be written:
\begin{align*}
\beta^{TWFE} = \frac{\E[\psi_j\beta_j]}{\E[\psi_j]},\qquad \psi_j=\tilde D_jD_j.
\end{align*}
Here $\tilde D_j$ is the residual from projecting $D_j$ on all fixed effects and controls while $\beta_j$ is the causal effect of treatment for observation $j$.  \citet{borusyakhull2024} call the $\psi_j$  \emph{ex post} weights because they depend on realized treatment assignments.\footnote{In contrast, in design-based specifications where treatment is conditionally as-good-as-randomly assigned, \cite{borusyakhull2024} show there is an equivalent representation of $\beta$ in terms of nonnegative \emph{ex ante} weights that average over treatment assignments---precluding sign reversals.} These weights can be negative because residualized treatment $\tilde D_j$ can be negative among treated observations, where $D_j=1$.  With sufficient treatment-effect heterogeneity, negative weights can place $\beta$ outside the convex hull of the effects. Event-study coefficients can likewise place negative weights on effects at the target relative time and can be contaminated by effects at other relative times \citep{sunabraham2021,goldsmithpinkhametal2024}; see  \citet{rothetal2023} and \citet{bakeretal2026} for more discussion.

The first modern estimator addresses these issues by separately estimating and averaging together conditional DiD contrasts with nonnegative weights, rather than relying on the implicit regression-weighting of TWFE. This DiD-averaging estimator follows \citet{callawaysantanna2021} while also being close to approaches in \cite{dechaisemartindhaultfoeuille2020} and  \cite{sunabraham2021}. For static specifications, the estimates take the form:
\begin{align*}
\hat\beta^{DA} = \sum_{g,t} \hat\omega_{g,t}(\bar{Y}_{g,t}-\bar{Y}_{g,b_g}-\widehat{\Delta Y}_{g,t}^0),
\end{align*}
where the $\hat\omega_{g,t}$ are convex weights, proportional to the number of treated observations in adoption cohort $g$ and post-treatment period $t$, $\bar{Y}_{g,t}$ and $\bar{Y}_{g,b_g}$ are sample-average outcomes for cohort $g$ in period $t$ and the last observed period before adoption, $b_g$, and $\widehat{\Delta Y}^0_{g,t}$ is an estimated counterfactual change in average outcomes without treatment.\footnote{I preserve each specification’s original reference-period normalization. When the reference interval includes some treated observations, the modern estimates subtract the estimated reference effect and therefore represent contrasts of treatment effects. Pretreatment contrasts in these cases are not pure placebo tests.} This counterfactual uses units that remain untreated through $t$, including never-treated units, adjusting linearly for the fixed characteristics $W_{u(j)}$. Under the no-anticipation and parallel-trends assumptions, the term in parentheses identifies the average treatment effect on the treated (ATT) in period $t$ for the cohort first treated in period $g$ and $\hat\beta^{DA}$ identifies the overall average. Event study estimates follow similarly, with cohort-average effects at different horizons first averaged with treated-count  weights then averaged together over pre- or post-treatment periods.

The second modern estimator follows the imputation approach of \citet*{borusyakjaravelspiess2024} (see also \citet{gardneretal2026}). The approach first estimates an untreated-outcome regression model by fitting equation \eqref{eq:twfe} via OLS without any treatment indicators and only on observations before treatment adoption, including any never-treated units. This returns fitted values $\hat{Y}_j(0)$ for all supported treated observations. The estimator is then:
\begin{align*}
\hat\beta^{IM} = |\mathcal{T}|^{-1}\sum_{j\in\mathcal{T}}(Y_j - \hat{Y}_j(0)),
\end{align*}
where $\mathcal{T}$ is the set of supported treated observations. Again, under the no-anticipation and parallel-trends assumptions, this estimates the ATT for supported observations. Post-treatment event-study summaries average the imputed effects at each horizon and then average bin coefficients equally.\footnote{Pretreatment summaries use either a separate untreated-sample regression with lead indicators or explicit-reference imputation contrasts; see Appendix \ref{app:screening}.} 

Alongside bias concerns from the potentially-negative weights in TWFE, one may also worry about the relatively higher variance of modern DiD estimators. Indeed, under constant treatment effects and the common identifying assumptions, the three treatment-effect estimands coincide; they can remain close when heterogeneity is small or weakly related to differences in their weights. But TWFE, by using more realized variation in the treatment, will tend to be more precise in these scenarios. I next probe these concerns empirically: first by comparing the $\hat\beta^{TWFE}$, $\hat\beta^{DA}$, and $\hat\beta^{IM}$ estimates themselves, and then via a hierarchical model that can separate estimation noise from true variation in the estimands.

\section{Results}\label{sec:results}

\subsection{Comparisons}
\begin{figure}[t!]
\centering
\caption{Modern Difference-in-Differences vs. Two-Way Fixed Effects}
\label{fig:twfe-baseline}
\includegraphics[width=\linewidth,trim=0pt 15pt 0pt 13pt,clip]{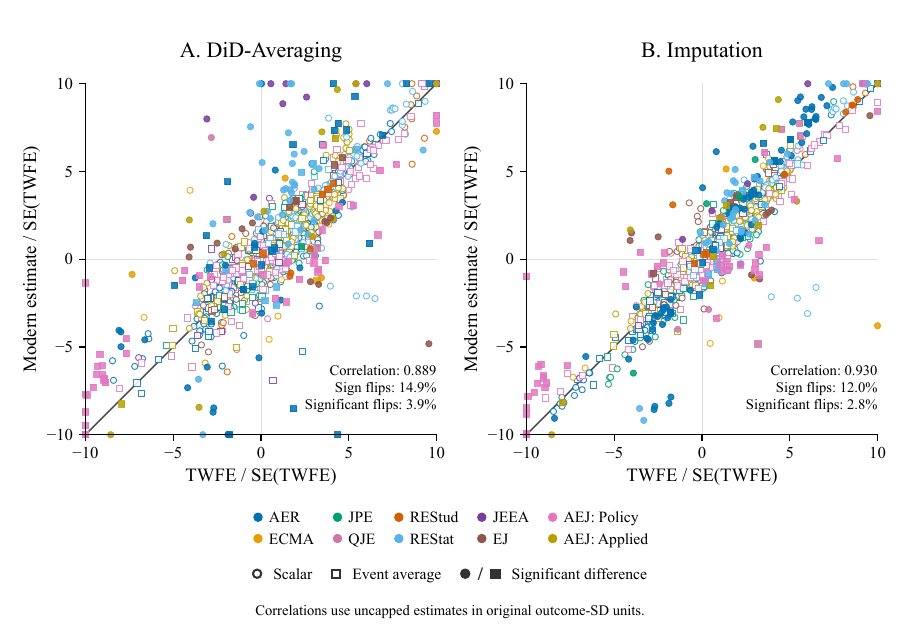}
\exhibitnotes{This figure compares DiD-averaging estimates (as in \citet{callawaysantanna2021}) and imputation estimates (as in  \citet*{borusyakjaravelspiess2024}) with two-way fixed-effects (TWFE) estimates. Each point is a scalar estimate (circle) or an event-study average (square), including separate pre- and post-treatment averages when available. The samples are described in Table \ref{tab:descriptive-statistics}. Both coordinates are divided by the TWFE standard error, and the diagonal denotes equal estimates.  Filled circles and squares indicate that the estimate difference is statistically significant at the 5\%  level. Coordinates are capped at $\pm10$; uncapped values, without standardizing by TWFE standard errors, are used for statistics. ``Sign flips'' reports the share of pairs with opposite signs; ``Significant flips'' reports the share with a sign flip and a significant difference.}
\end{figure}

TWFE estimates are highly correlated with modern DiD estimates in this sample. The sample correlation is 0.89 for DiD-averaging and 0.93 for imputation. Correspondingly, sign reversals are relatively infrequent---and especially so  with statistically significant estimator differences. Around 15.0\% of TWFE estimates change sign with DiD averaging while 3.8\% do so with a significant difference. These shares are lower with imputation, at 12.0\% and 2.7\%.  Figure \ref{fig:twfe-baseline} plots the full distribution of estimates.\footnote{For comparability with \cite{chiuetal2026}, points on this graph are normalized by the TWFE standard error. All statistics are computed from the underlying outcome-standardized estimates. Appendix Figure \ref{fig:twfe-baseline-no-pre} shows a version of the plot excluding pre-treatment comparisons from event studies. Sign reversal rates are lower in this sample: 12.3\% for DiD-averaging and 8.2\% for imputation. }

One likely reason for the infrequency of sign reversals is the relatively small footprint of negative \emph{ex post} weights. Among static TWFE specifications, the median fraction of treated observation rows with negative ex post weights is approximately 8\%, while the median share of absolute effect weight carried by negative weights is only 1.4\%  (Appendix Figure  \ref{fig:negative-weights} shows full distributions).\footnote{Appendix Figure \ref{fig:static-negative-weight-gaps} further shows that estimate sign reversals and large movements in TWFE vs. modern DiD estimates are concentrated in specifications with high negative weight shares.} Hence, while negative weights are theoretically important, their mass is limited in the typical specification.\footnote{Failures of the no-anticipation and parallel-trends assumptions may also drive differences in estimands, along with small differences in group and time aggregation; estimation error can further drive some observed reversals in Figure \ref{fig:twfe-baseline}. Appendix Figure~\ref{fig:csa-vs-bjs-twfe-se} compares the DiD-averaging and imputation estimators directly; they are highly but not perfectly correlated, suggesting these other factors also play a role.}

The Figure \ref{fig:twfe-baseline} scatterplots also show substantially more dispersion in the DiD-averaging estimates than in the imputation estimates, which are themselves more dispersed than the TWFE estimates. These suggest increased noise, assuming similar estimands, and indeed DiD-averaging and imputation standard errors are on average 40\% and 9\% larger than corresponding TWFE standard errors. The findings thus illustrate the tradeoff between robustness to the potential bias from TWFE weighting and the relative precision of OLS. The relative precision of imputation vs. DiD-averaging also helps explain its lower rate of sign reversals in Figure \ref{fig:twfe-baseline}. 

\subsection{Correlates}
\begin{figure}[t!]
\centering
\caption{Correlates of Sign Reversals}
\label{fig:sign-reversal-predictors}
\includegraphics[width=\linewidth,trim=0pt 10pt 0pt 7pt,clip]{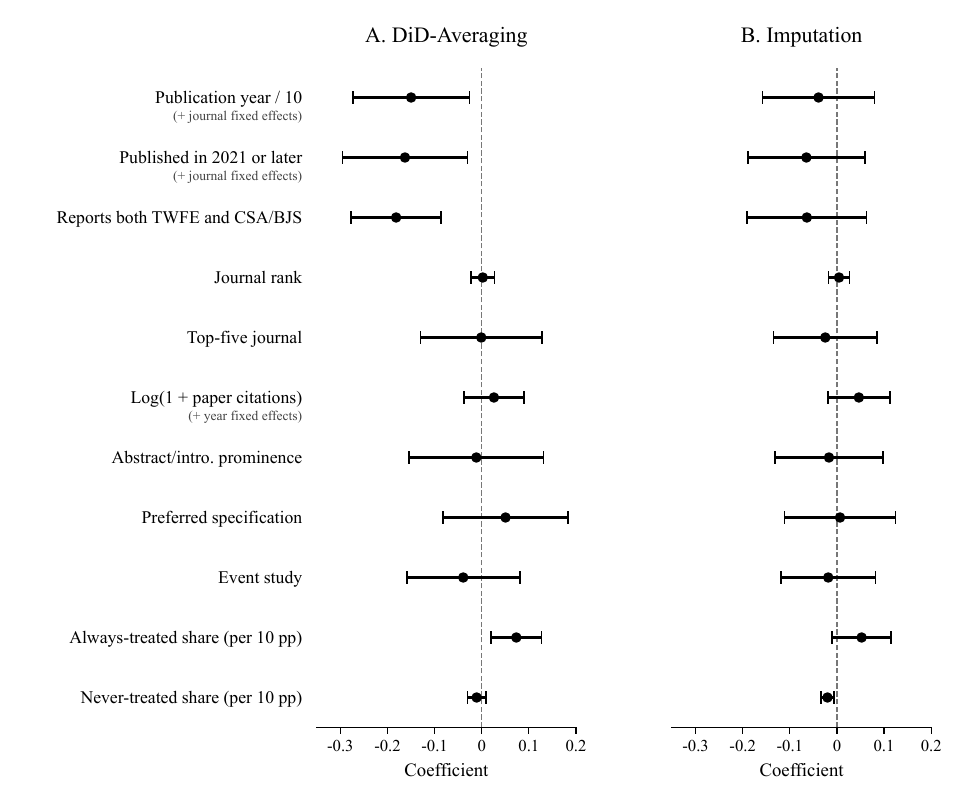}
\exhibitnotes{Points are coefficients from separate weighted regressions of sign reversal on each predictor. The first two rows include journal fixed effects; the citation row includes publication-year fixed effects. All other rows have no additional controls. Whiskers are 95\% intervals using paper-clustered standard errors. Inverse-squared TWFE standard-error weights sum to one within paper and estimator after each restriction. The sample is as in Figure~\ref{fig:twfe-baseline}: 1,205 DiD-averaging observations from 90 papers and 1,254 imputation observations from 97 papers. The preferred specification regression instead uses 935 observations from 67 papers in panel A and 969 from 73 in panel B. Journal ranks follow RePEc order, 1--10 from highest to lowest standing. Always-treated and never-treated share coefficients represent associations per 10 percentage points. See Table \ref{tab:descriptive-statistics} for other variable definitions.}
\end{figure}

TWFE sign reversals have declined over time, particularly after the publication of influential econometrics papers establishing the concern. Figure \ref{fig:sign-reversal-predictors} plots coefficients from regressing sign flip indicators on various specification and paper characteristics; we weight inversely  by the TWFE estimate variance (standard error squared), normalized by paper such that each paper receives equal weight. The first two rows show that later publication years are associated with a statistically significantly lower rate of a sign reversal: when moving from TWFE to a DiD-averaging specification, there is a decline of around 15 percentage points per decade and an overall average decline of 16 percentage points for papers published in 2021 and later.\footnote{For the specifications with an indicator for publication in 2021 or later, intercept estimates are 0.20 for DiD-averaging and 0.09 for imputation, with standard errors of 0.05 and 0.04, respectively.} Trends for imputation are similarly negative, but not statistically significant.\footnote{Publication-time specifications control for journal fixed effects to address the changing composition.}

Over-time declines are consistent with increased awareness of---and endogenous responses to---the TWFE negative weight concern. Another finding consistent with this hypothesis is shown in the third row of Figure \ref{fig:sign-reversal-predictors}: papers that report both TWFE and a modern DiD estimator (either from \cite{callawaysantanna2021} or \cite{borusyakjaravelspiess2024}) see a significant 18 percentage point lower probability of sign reversals from DiD-averaging. Again, for imputation this correlation is also negative but insignificant. Of course we cannot rule out the possibility that sign reversals decline for some other reason, such as the increased number of other robustness checks (e.g. pre-trend or subsample analyses).

Cross-sectionally, while sign reversal rates vary across journals, there is no statistically significant association with journal rank. The next two rows of Figure \ref{fig:sign-reversal-predictors} show statistically insignificant coefficients from regression on RePEc journal rankings and an indicator for top-five status. Similarly, the relationship with paper citations is insignificant.\footnote{This specification controls for year fixed effects to address different horizons for accumulating citations.} 

Furthermore, there is no statistically significant association between sign reversals and whether specifications appear to be more prominent or preferred. The next two rows of Figure \ref{fig:sign-reversal-predictors} show statistically insignificant coefficients on both the indicator for a result being mentioned in the paper's abstract or introduction and our constructed indicator for a preferred specification. The following row shows that event studies are somewhat less likely to exhibit sign reversals for DiD-averaging, though this coefficient is statistically insignificant.

The final two rows of Figure \ref{fig:sign-reversal-predictors} establish two strong predictors of TWFE sign reversals which are consistent with the econometric theory and easily computed in a specification: the shares of units which are always treated or never treated. Always-treated units generally increase TWFE's exposure to ``forbidden comparisons,'' in which already-treated units serve as controls for units newly adopting treatment \citep{goodmanbacon2021,borusyakjaravelspiess2024}. Empirically, I find that a ten percentage point increase in a specification's always-treated share predicts a statistically significant 7.3 percentage point increase in the rate of sign reversals for DiD-averaging; for imputation, this coefficient is a similar (though not significant) 5.2 percentage points. Never treated units provide ``clean controls,'' so a larger share of them can reduce exposure to forbidden comparisons. Here for imputation I find a statistically significant reduction of 2.0 percentage points in sign reversals per 10 percentage point increase in the never-treated share; for DiD averaging, this coefficient is also negative but insignificant.

\subsection{Hierarchical Model}

The final set of analyses uses a hierarchical model to characterize the distribution of TWFE and modern DiD estimands, separating variation in estimands from sampling error. The model builds on ones proposed in \citet{chen2026close} and \citet{walters2024empirical} in allowing the first and second moments of estimands to depend on estimator precision; in the baseline model, residual variation is modeled as Gaussian.

Specifically, the model uses the usual asymptotic approximation to specify pairs of TWFE and modern DiD estimates $\hat\theta_k\in\mathbb{R}^2$ for each comparison point $k$ as bivariate normal around those specifications' estimands $\theta_k$, with variances given by squared standard errors:
\begin{align*}
\hat\theta_k \mid \theta_k, S_k \sim N(\theta_k, S_k),
\end{align*}
where $S_k$ denotes the estimated variance-covariance matrix for $\hat\theta_k$, including the covariance between the two estimators.\footnote{Covariances are obtained by a weighted (Bayesian) bootstrap with 999 replications.} The Gaussian baseline model then specifies:
\begin{align*}
\theta_k \mid S_k &\sim N\left(\begin{pmatrix}
\mu_{kT} \\
\mu_{kM}
\end{pmatrix},\begin{pmatrix}
\tau_{kT}^{2} & \rho\tau_{kT}\tau_{kM}\\
\rho\tau_{kT}\tau_{kM} & \tau_{kM}^{2}
\end{pmatrix}\right) ,
\end{align*}
where $\mu_{ke}=a_e+b_eQ_{ke}$ and $\tau_{ke}=\exp(c_e+d_eQ_{ke})$ for $e\in\{T,M\}$ (TWFE or modern DiD) and where $Q_{ke}$ denotes the demeaned log standard error of specification $k$ with estimator $e$. Thus each estimand's marginal mean and standard deviation depend on its estimator precision. The model has a total of nine parameters, $(a_T,a_M,b_T,b_M,c_T,c_M,d_T,d_M,\rho)$ for each set of comparisons. I estimate these parameters by maximum likelihood, first profiling out the mean parameters by generalized least squares, with inference clustered by paper. This is a bivariate version of the Gaussian precision-dependent model in \citet{walters2024empirical}. I also check sensitivity to the analogous nonparametric-prior model as in \citet{chen2026close}.\footnote{Estimation of this model follows the procedure \citet{chen2026close} outlines for his univariate CLOSE model. I use local-linear smoothing against log standard errors with IMSE-DPI bandwidth selection to estimate precision dependence, then fit the prior distribution by nonparametric maximum likelihood estimation. I fit the prior for pre-treatment points from event studies with a separate nonparametric prior.}

\begin{table}[t!]
\centering
\caption{Hierarchical Model Estimates}
\label{tab:gaussian-priors}
\small
\begin{tabular}{@{}p{0.56\linewidth}cc@{}}
\toprule
 & DiD-Averaging & Imputation \\
 & (1) & (2) \\
\midrule
Standard deviation of TWFE estimands & 0.195 & 0.185 \\
 & (0.016) & (0.016) \\
\addlinespace[5pt]
Standard deviation of modern-DiD estimands & 0.215 & 0.213 \\
 & (0.022) & (0.022) \\
\addlinespace[5pt]
Correlation of TWFE and modern-DiD estimands & 0.911 & 0.971 \\
 & (0.023) & (0.006) \\
\addlinespace[5pt]
Prior probability of a sign reversal & 0.118 & 0.060 \\
 & (0.018) & (0.008) \\
\addlinespace[5pt]
Significant wrong-signed TWFE probability & 0.024 & 0.007 \\
 & (0.007) & (0.002) \\
\midrule
Observations & 1,205 & 1,254 \\
Papers & 90 & 97 \\
\bottomrule
\end{tabular}
\exhibitnotes{This table reports estimates from the Gaussian hierarchical model described in the text. Estimates come from averaging priors over observed standard errors, giving equal weight to each retained comparison point. Standard deviations are in  regression-sample outcome standard deviations. A sign reversal is $\theta_T\theta_M<0$. A significant wrong-signed TWFE estimate is $\widehat\theta_T\theta_M<0$ and $|\widehat\theta_T|>1.96s_T$, where $\widehat\theta_T\mid\theta_T,s_T\sim N(\theta_T,s_T^2)$. Parentheses report delta-method standard errors clustered by paper, incorporating uncertainty in the fitted parameters and empirical precision mixture. The sample is the same as in Figure~\ref{fig:twfe-baseline}. }
\end{table}

Model estimates show that the high correlation between TWFE and modern DiD estimators in Figure \ref{fig:twfe-baseline} is not driven by correlated estimation error. Table \ref{tab:gaussian-priors} reports estimand correlations of 0.91 for DiD-averaging and 0.97 for imputation.  TWFE and modern DiD estimands have similar standard deviations of around 0.2 (in units of observation-level outcome standard deviations) with modern estimands being somewhat more variable. This suggests that the ex post weights of TWFE may somewhat understate the dispersion of unconditional-average causal effects.  

Estimand sign reversals are rarer than the estimate reversal rates in Figure \ref{fig:twfe-baseline} suggest. Table \ref{tab:gaussian-priors}  reports that around 11.8\% of TWFE and modern DiD estimand pairs have opposite signs, as opposed to 14.9\% of estimate pairs. For imputation, these rates are 6.0\% for estimands versus 12.0\% for estimates. Naturally, some estimate sign flips reflect finite-sample noise which the model can account for. The model also allows us to infer how often TWFE would lead to an incorrect judgement on the modern DiD estimand's sign: i.e., the probability of a researcher finding a significantly positive (or negative) TWFE estimate when the modern DiD estimand is negative (or positive). These rates are 2.4\% for DiD averaging and 0.7\% for imputation, reflecting the high correlation in estimands.\footnote{For context, the probability of obtaining a significant wrong-signed estimate with either modern DiD estimator is 0.27\% (DiD-averaging) and 0.24\% (imputation).} In other words, in this sample, sign reversals do not tend to move the TWFE estimand far below (or above) zero relative to the estimator's standard error.\footnote{Appendix Table \ref{tab:gaussian-priors-no-pre} shows similar estimates after excluding pre-treatment comparisons from event studies.} 

Importantly, these qualitative conclusions persist under the alternative nonparametric specification. Nonparametric estimates of estimand joint distributions, plotted in Figure \ref{fig:precision-dependent-priors}, give latent sign-reversal rates of 12.0\% for DiD-averaging and 10.0\% for imputation, compared with 11.8\% and 6.0\% in the Gaussian model. Probabilities of significant sign errors are 2.4\% and 1.8\%, compared with 2.4\% and 0.7\% in the Gaussian model.

\section{Conclusion}\label{sec:conclusion}

The empirical relevance of theoretical econometric concerns is rarely evaluated systematically. This paper shows how modern AI tools can be used to reanalyze a large number of published economic specifications and assess a recent concern that has reshaped how the profession thinks about and approaches staggered adoption designs. Findings suggest that negative weights, while potentially a serious problem for TWFE, are unlikely to dramatically affect empirical conclusions across the large number of replicated specifications. Future work could apply similar large-scale meta-analyses to other high-profile concerns, such as weak instruments, standard error clustering, or insufficiently flexible regression adjustment.

Importantly, this paper does not take a stance on whether the estimated sign reversal rates or estimand correlations are ``low enough'' to dismiss concerns over TWFE estimation. Modern DiD estimators are easily run in most circumstances, and checking their empirical relevance is likely useful for any given staggered-adoption application. Instead, this paper gives a systematic empirical foundation to ongoing discussions about relative tradeoffs between TWFE and alternative estimators, which can of course have their own drawbacks (such as imprecision). The findings may help shape further research and practical guidance for navigating the many potential pitfalls of TWFE-based causal inference---including parallel trend violations and standard error misspecification, alongside negative weights.

\singlespacing

\begingroup
\singlespacing
\setlength{\bibsep}{6pt plus 1pt}
\interlinepenalty=10000
\section*{References}
\addcontentsline{toc}{section}{References}
\renewcommand{\bibsection}{}
\bibliographystyle{aer}
\bibliography{indifference_refs}
\endgroup
\clearpage

\appendix
\counterwithin{figure}{section}
\counterwithin{table}{section}
\renewcommand{\thefigure}{\thesection\arabic{figure}}
\renewcommand{\thetable}{\thesection\arabic{table}}
\renewcommand{\theHfigure}{appendix.\thesection.\arabic{figure}}
\renewcommand{\theHtable}{appendix.\thesection.\arabic{table}}
\section{Additional Exhibits}
\singlespacing

\begin{figure}[!ht]
\centering
\caption{Modern Difference-in-Differences vs. Two-Way Fixed Effects, \\ Excluding Pre-Treatment Comparisons}
\label{fig:twfe-baseline-no-pre}
\includegraphics[width=\linewidth,trim=0pt 15pt 0pt 13pt,clip]{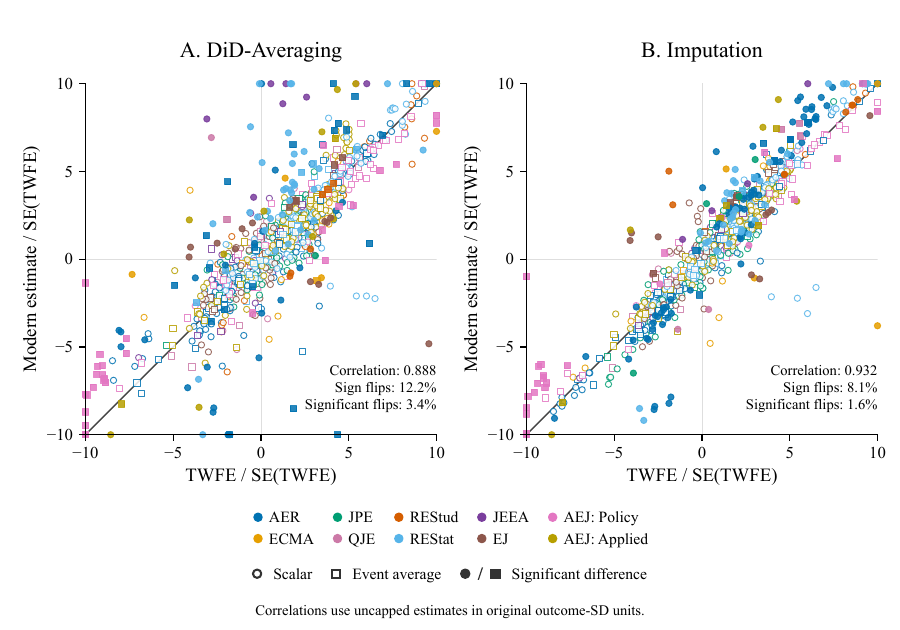}
\exhibitnotes{This figure compares DiD-averaging estimates (as in \citet{callawaysantanna2021}) and imputation estimates (as in  \citet*{borusyakjaravelspiess2024}) with two-way fixed-effects (TWFE) estimates. Each point is a scalar estimate (circle) or a post-treatment event-study average (square). The samples are those in Figure~\ref{fig:twfe-baseline} after excluding all pre-treatment comparisons: 988 DiD-averaging observations from 89 papers and 1,019 imputation observations from 93 papers. Both coordinates are divided by the TWFE standard error, and the diagonal denotes equal estimates.  Filled circles and squares indicate that the estimate difference is statistically significant at the 5\%  level. Coordinates are capped at $\pm10$; uncapped values, without standardizing by TWFE standard errors, are used for statistics. ``Sign flips'' reports the share of pairs with opposite signs; ``Significant flips'' reports the share with a sign flip and a significant difference.}
\end{figure}
\clearpage

\begin{figure}[!ht]
\centering
\caption{DiD-Averaging vs. Imputation}
\label{fig:csa-vs-bjs-twfe-se}
\includegraphics[width=0.8\linewidth]{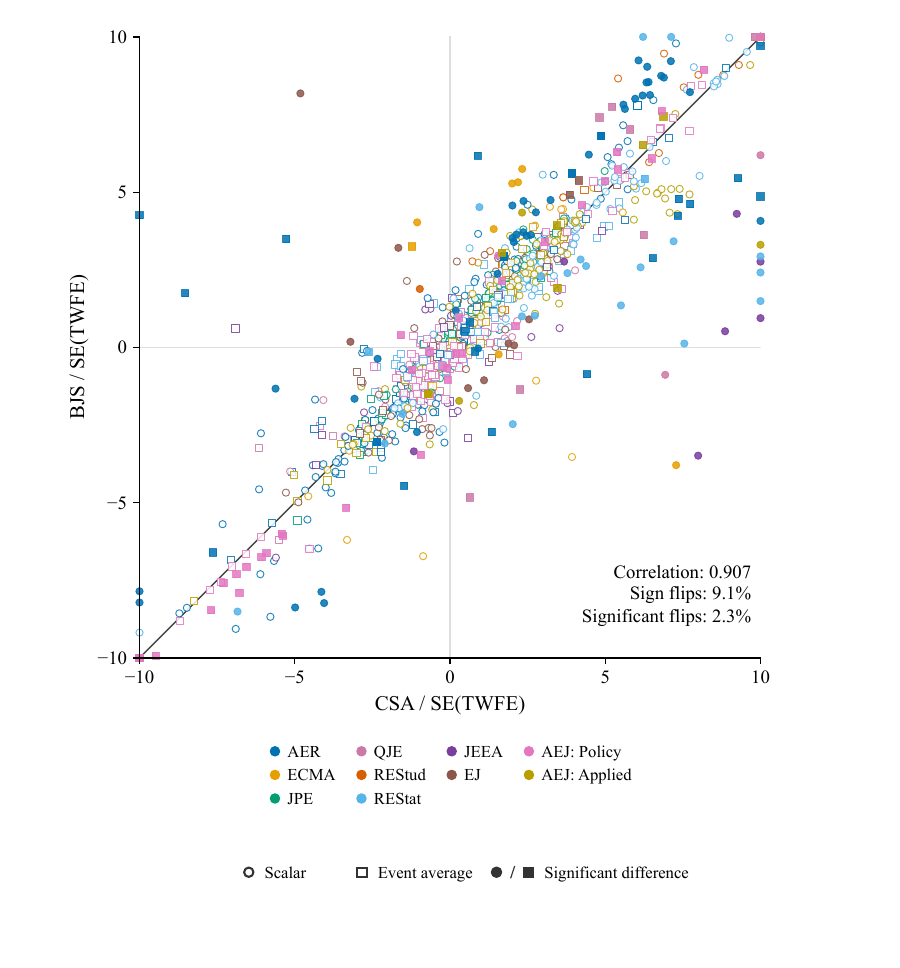}
\exhibitnotes{This figure compares DiD-averaging and imputation estimates for the 1,137 common comparison points from 88 papers passing both estimators' leverage screens. Each point is a scalar estimate (circle) or an event-study average (square), including separate pre- and post-treatment averages when available. Both coordinates are divided by the corresponding TWFE standard error, and the diagonal denotes equal estimates. Filled circles and squares indicate that the estimate difference is statistically significant at the 5\% level. Coordinates are capped at $\pm10$; uncapped values are used for statistics. ``Sign flips'' reports the share of pairs with opposite signs; ``Significant flips'' reports the share with a sign flip and a significant paired difference.}
\end{figure}
\clearpage

\begin{figure}[!ht]
\centering
\caption{Negative Weights in Two-Way Fixed-Effects Estimates}
\label{fig:negative-weights}
\makebox[\linewidth][c]{\includegraphics[width=1\linewidth,trim=0pt 14pt 0pt 7pt,clip]{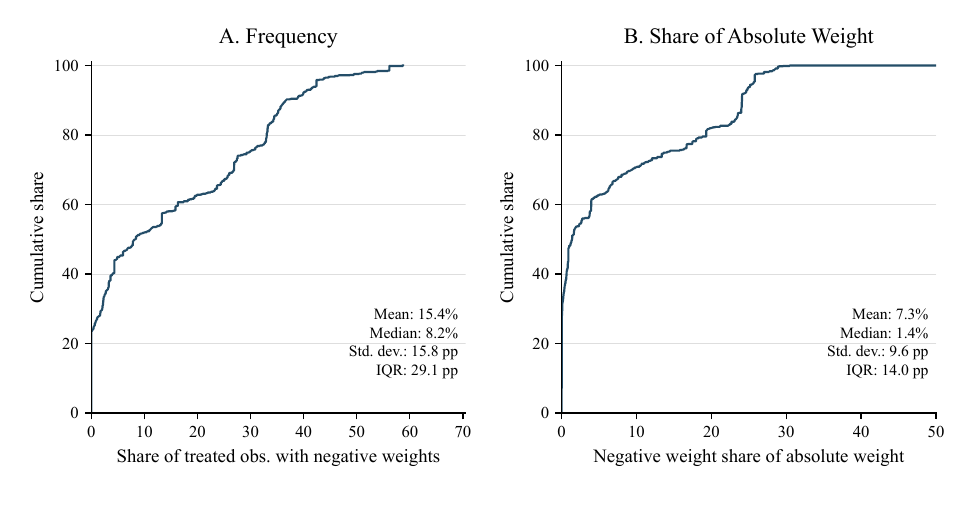}}
\exhibitnotes{This figure summarizes the distribution of negative \emph{ex post} weights $\psi_j$ across the 1,038 static TWFE specifications from 91 papers in the replication sample. Event-study contrasts are excluded. The sample includes 53 constructed static TWFE counterparts for specifications originally reporting modern DiD estimators. Panel A shows the cumulative distribution function for the fraction of treated observation rows with negative weights, including zero-weight rows in the denominator. Panel B shows the distribution for the share of absolute effect weight carried by negative weights, $-\sum_j\min(\psi_j,0)/\sum_j|\psi_j|$. Annotations report the equally weighted specification-level mean, median, standard deviation, and interquartile range (IQR).}
\end{figure}
\clearpage

\begin{figure}[!ht]
\centering
\caption{Negative Weights and Estimator Differences}
\label{fig:static-negative-weight-gaps}
\includegraphics[width=\linewidth]{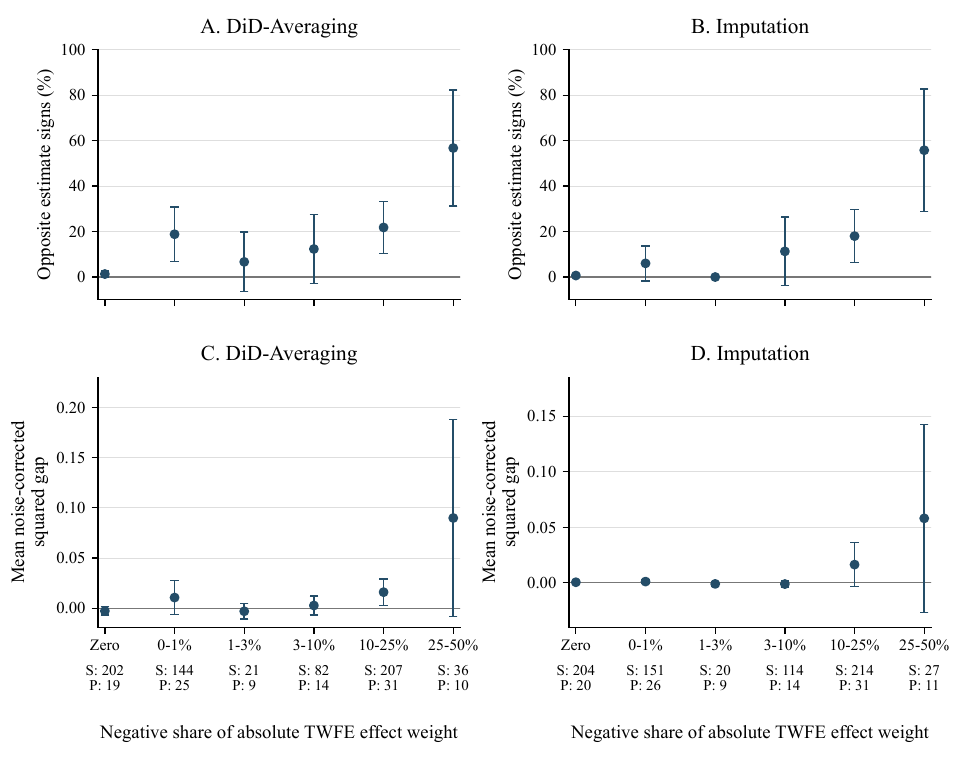}
\exhibitnotes{This figure relates negative TWFE weight shares to differences between TWFE and modern DiD estimates for static specifications. Panels A and B show the share of estimate pairs with opposite signs for DiD-averaging and imputation, respectively. Panels C and D show mean squared estimate differences after subtracting the estimated variance of each paired difference, including the covariance between estimators. Both estimates in a pair are normalized by the original regression-sample outcome standard deviation. Negative weight shares are defined as in Figure~\ref{fig:negative-weights}. Within each bin, papers receive equal weight and specifications receive equal weight within papers. Bars show 95\% confidence intervals, clustered by paper. The S and P numbers report the specification and paper counts in each bin.}
\end{figure}
\clearpage

\begin{figure}[p]
\centering
\caption{Estimand Distributions, Parametric vs. Nonparametric }
\label{fig:precision-dependent-priors}
\includegraphics[width=0.97\linewidth,trim=0pt 0pt 0pt 2pt,clip]{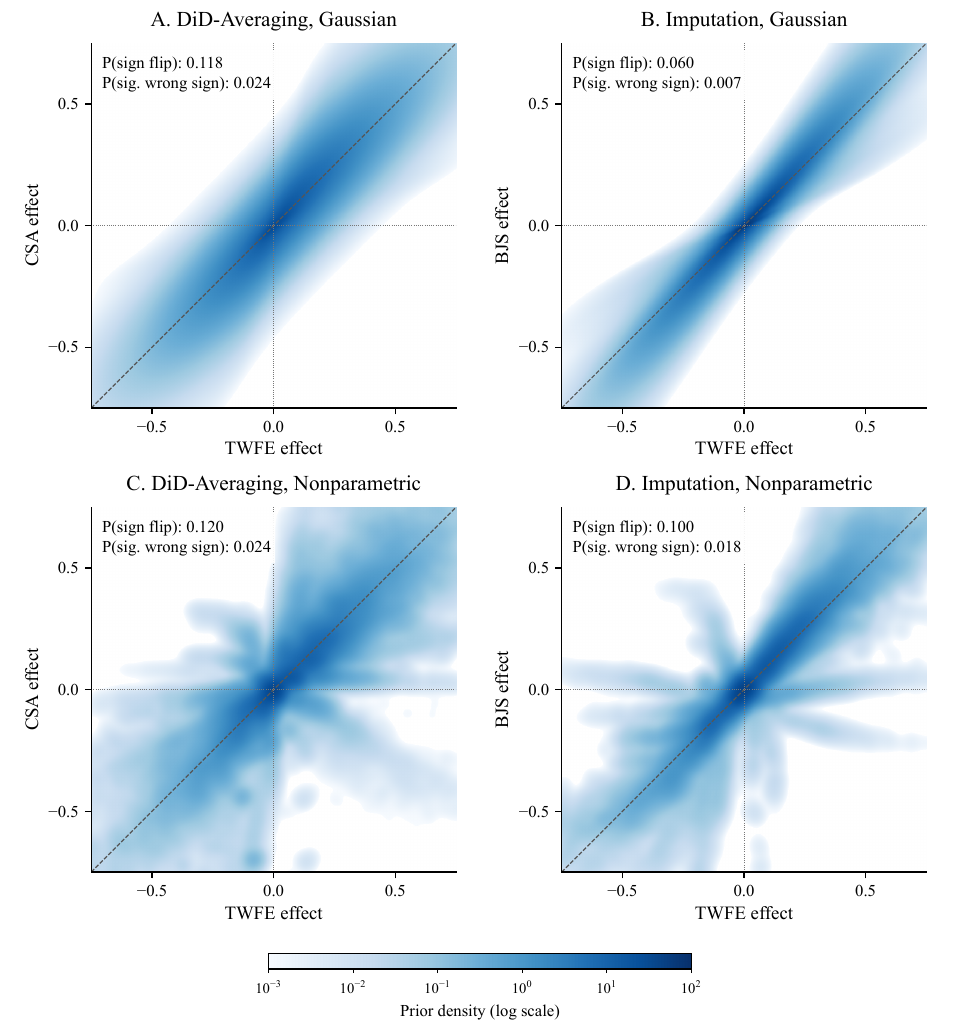}
\exhibitnotes{This figure shows estimated joint distributions of TWFE and modern DiD estimands. Panels A and B use the baseline Gaussian model described in the text while panels C and D use a nonparametric model for the prior distribution, with separate
 precision-dependent means, scales, and mixing distributions for pretreatment and static/post-treatment comparisons.  Distributions are averaged over empirical standard-error pairs. The nonparametric density is smoothed with an isotropic Gaussian kernel of standard deviation 0.025 per coordinate. Annotations report full-prior sign-flip and significant wrong-signed TWFE probabilities using the definitions in Table~\ref{tab:gaussian-priors}. Estimands are plotted in outcome-standard-deviation units. All panels share the logarithmic density scale; white denotes density below 0.001.  The sample is the same as in Figure~\ref{fig:twfe-baseline}.}
\end{figure}
\clearpage

\begin{table}[p]
\centering
\caption{Papers and Contributions to the Analysis}
\label{tab:paper-inventory}
\begingroup
\fontsize{7.7}{8.8}\selectfont
\setlength{\tabcolsep}{1.5pt}
\renewcommand{\arraystretch}{0.965}
\makebox[\linewidth][c]{%
\begin{tabular}{@{}p{0.425\linewidth}rrr@{\hspace{12pt}}p{0.37\linewidth}rrr@{}}
\toprule
Paper & Specs. & CSA & BJS & Paper & Specs. & CSA & BJS \\
\midrule
\raggedright Abeberese et~al. (REStat '23) & 7 & 8 & 7 & \raggedright Dippel and Heblich (AER '21) & 2 & 2 & 2 \\
\raggedright Acemoglu and Azar (ECMA '20) & 3 & 3 & 3 & \raggedright Donaldson (AER '18) & 1 & 1 & 1 \\
\raggedright Ager et~al. (REStat '20) & 1 & 2 & 2 & \raggedright Egedesø et~al. (EJ '20) & 20 & 22 & 22 \\
\raggedright Aghion et~al. (AER '08) & 2 & 2 & 2 & \raggedright Esposito et~al. (AER '23) & 8 & 8 & 8 \\
\raggedright Agrawal and Goldfarb (AER '08) & 1 & 1 & 1 & \raggedright Falcone (EJ '26) & 7 & 12 & 12 \\
\raggedright Akerlof et~al. (JEEA '26) & 2 & 2 & 1 & \raggedright Fasani et~al. (JEEA '21) & 2 & 0 & 0 \\
\raggedright Aman-Rana and Minaudier (AER '26) & 14 & 16 & 16 & \raggedright Fetzer (EJ '22) & 3 & 1 & 1 \\
\raggedright Andersson et~al. (REStat '23) & 39 & 39 & 39 & \raggedright Fetzer et~al. (REStat '21) & 4 & 4 & 4 \\
\raggedright Andrews (AEJ: Policy '23) & 9 & 6 & 5 & \raggedright Frakes (AER '13) & 3 & 0 & 0 \\
\raggedright Ash et~al. (AEJ: Policy '25) & 1 & 1 & 1 & \raggedright Funke et~al. (AER '23) & 2 & 0 & 0 \\
\raggedright Askarov et~al. (JEEA '23) & 4 & 0 & 0 & \raggedright Galiani et~al. (JPE '05) & 2 & 2 & 2 \\
\raggedright Avis et~al. (JPE '18) & 3 & 3 & 3 & \raggedright Gibson (REStat '19) & 44 & 15 & 13 \\
\raggedright Bahaj and Reis (REStud '26) & 16 & 16 & 1 & \raggedright Gibson and McKenzie (REStat '14) & 60 & 60 & 60 \\
\raggedright Bailey (AEJ: App '12) & 11 & 11 & 12 & \raggedright Giorcelli and Moser (JPE '20) & 4 & 0 & 4 \\
\raggedright Bailey and Goodman-Bacon (AER '15) & 2 & 4 & 4 & \raggedright Gokmen et~al. (EJ '21) & 36 & 36 & 25 \\
\raggedright Bancalari (REStat '26) & 32 & 52 & 52 & \raggedright González et~al. (EJ '24) & 12 & 12 & 12 \\
\raggedright Banerjee et~al. (AEJ: App '20) & 1 & 1 & 1 & \raggedright Gonçalves et~al. (AER '26) & 15 & 0 & 0 \\
\raggedright Baum-Snow and Lutz (AER '11) & 55 & 21 & 25 & \raggedright Hamory et~al. (JEEA '21) & 14 & 11 & 14 \\
\raggedright Bazzi et~al. (REStud '26) & 18 & 24 & 24 & \raggedright Hansman et~al. (JPE '20) & 7 & 0 & 5 \\
\raggedright Becker et~al. (JEEA '25) & 3 & 0 & 1 & \raggedright He and Wang (AEJ: App '17) & 12 & 12 & 15 \\
\raggedright Beg (JEEA '22) & 4 & 6 & 6 & \raggedright Hjort and Poulsen (AER '19) & 3 & 2 & 2 \\
\raggedright Bellégo and Drouard (AEJ: Policy '24) & 21 & 42 & 21 & \raggedright Hodler and Raschky (QJE '14) & 12 & 14 & 14 \\
\raggedright Benzell and Cooke (AEJ: App '21) & 2 & 2 & 0 & \raggedright Hollingsworth et~al. (AER '24) & 81 & 96 & 96 \\
\raggedright Bertrand et~al. (QJE '21) & 1 & 1 & 1 & \raggedright Iyer et~al. (AEJ: App '12) & 9 & 0 & 0 \\
\raggedright Bhalotra et~al. (JEEA '23) & 25 & 21 & 21 & \raggedright Janssen and Zhang (AER '23) & 2 & 2 & 2 \\
\raggedright Bindler and Hjalmarsson (AEJ: Policy '18) & 10 & 0 & 0 & \raggedright Jensen and Miller (AER '18) & 7 & 7 & 7 \\
\raggedright Boomhower (AER '19) & 1 & 1 & 1 & \raggedright Jia et~al. (JEEA '15) & 10 & 0 & 10 \\
\raggedright Borcan et~al. (AEJ: Policy '17) & 5 & 5 & 5 & \raggedright Johnson and Noguera (REStat '17) & 6 & 6 & 6 \\
\raggedright Bostwick et~al. (AEJ: Policy '22) & 2 & 0 & 1 & \raggedright Johnson et~al. (REStat '24) & 4 & 0 & 0 \\
\raggedright Braga and Malkova (JEEA '24) & 6 & 0 & 0 & \raggedright Kreindler et~al. (AER '26) & 25 & 26 & 24 \\
\raggedright Braguinsky et~al. (AER '15) & 8 & 0 & 13 & \raggedright Kremer et~al. (QJE '11) & 1 & 1 & 1 \\
\raggedright Brehm and Saavedra (EJ '25) & 13 & 4 & 6 & \raggedright La Ferrara et~al. (AEJ: App '12) & 3 & 3 & 3 \\
\raggedright Buchmueller and Carey (AEJ: Policy '18) & 38 & 0 & 0 & \raggedright Lafortune et~al. (AEJ: App '18) & 46 & 43 & 41 \\
\raggedright Calamunci and Lonsky (EJ '25) & 3 & 3 & 3 & \raggedright Le Barbanchon and Sauvagnat (JEEA '22) & 2 & 2 & 2 \\
\raggedright Cantoni and Pons (QJE '21) & 14 & 0 & 0 & \raggedright Lehmann-Hasemeyer and Streb (AER '16) & 4 & 8 & 8 \\
\raggedright Cantoni et~al. (ECMA '24) & 14 & 10 & 14 & \raggedright Lewis-Faupel et~al. (AEJ: Policy '16) & 2 & 0 & 0 \\
\raggedright Cantoni et~al. (JPE '17) & 103 & 103 & 103 & \raggedright Martinez-Bravo (AEJ: App '17) & 4 & 4 & 4 \\
\raggedright Carrillo et~al. (AEJ: Policy '23) & 1 & 2 & 2 & \raggedright Martinez-Bravo et~al. (AER '22) & 5 & 4 & 4 \\
\raggedright Cervellati et~al. (EJ '17) & 2 & 2 & 2 & \raggedright Mastrorocco and Teso (REStud '26) & 1 & 1 & 1 \\
\raggedright Chen and Lan (AEJ: App '17) & 1 & 1 & 1 & \raggedright Moser et~al. (AER '14) & 5 & 5 & 5 \\
\raggedright Chen et~al. (EJ '22) & 1 & 1 & 1 & \raggedright Muñoz (QJE '24) & 3 & 4 & 4 \\
\raggedright Chiovelli et~al. (ECMA '25) & 11 & 11 & 11 & \raggedright Nagengast and Yotov (AEJ: App '25) & 2 & 2 & 2 \\
\raggedright Chirakijja (REStat '24) & 9 & 12 & 12 & \raggedright Noack and Costello (JPE '24) & 1 & 2 & 2 \\
\raggedright Chu et~al. (AEJ: App '21) & 2 & 1 & 1 & \raggedright Norris and Xiong (EJ '23) & 9 & 9 & 9 \\
\raggedright Cicala (AER '15) & 4 & 0 & 8 & \raggedright Priks (EJ '15) & 15 & 1 & 0 \\
\raggedright Cohen et~al. (JPE '11) & 1 & 0 & 1 & \raggedright Rexer (AER '25) & 90 & 37 & 61 \\
\raggedright Cohen et~al. (JPE '17) & 16 & 0 & 0 & \raggedright Shin (REStat '25) & 1 & 2 & 2 \\
\raggedright Colonnelli and Prem (REStud '22) & 5 & 5 & 5 & \raggedright Snyder and Welch (JPE '17) & 21 & 0 & 16 \\
\raggedright Conley and Taber (REStat '11) & 1 & 0 & 0 & \raggedright Spenkuch et~al. (ECMA '23) & 3 & 3 & 3 \\
\raggedright Cook and Shah (REStat '22) & 4 & 4 & 4 & \raggedright Starc and Wollmann (AER '25) & 4 & 8 & 8 \\
\raggedright Coube et~al. (QJE '26) & 12 & 18 & 18 & \raggedright Stevenson and Wolfers (QJE '06) & 8 & 8 & 8 \\
\raggedright Couttenier et~al. (JEEA '17) & 1 & 0 & 0 & \raggedright Sun and Yannelis (REStat '16) & 2 & 2 & 2 \\
\raggedright Cunningham and Goodman-Bacon (AEJ: App '25) & 36 & 47 & 44 & \raggedright Sánchez de la Sierra et~al. (AER '24) & 32 & 0 & 0 \\
\raggedright Dahis and Szerman (AEJ: App '26) & 5 & 0 & 0 & \raggedright Weill (AEJ: Policy '26) & 48 & 96 & 96 \\
\raggedright Dal Bó et~al. (REStud '09) & 2 & 1 & 2 & \raggedright Wolfers (AER '06) & 12 & 2 & 11 \\
\raggedright Davis and Wolfram (AEJ: App '12) & 37 & 35 & 36 & \raggedright Yang et~al. (AEJ: Policy '24) & 1 & 1 & 1 \\
\raggedright de Janvry et~al. (AER '15) & 9 & 9 & 9 & \raggedright Yousaf (JEEA '21) & 1 & 1 & 1 \\
\raggedright Deryugina (AEJ: Policy '17) & 28 & 41 & 41 & \raggedright Zussman (AEJ: App '23) & 5 & 5 & 5 \\
\midrule
\multicolumn{5}{@{}l}{Total (116 papers)} & 1,430 & 1,205 & 1,254 \\
\bottomrule
\end{tabular}}
\endgroup
\exhibitnotes{This table lists the 116 papers with replicated specifications. Specs. counts all 1,430 specifications, including those that contribute no comparison plotted in Figure~\ref{fig:twfe-baseline}; CSA and BJS count retained DiD-averaging and imputation comparison points plotted in that figure, respectively.}
\end{table}
\clearpage

\begin{table}[p]
\centering
\caption{Hierarchical Model Estimates, Excluding Pre-Treatment Comparisons}
\label{tab:gaussian-priors-no-pre}
\small
\begin{tabular}{@{}p{0.56\linewidth}cc@{}}
\toprule
 & DiD-Averaging & Imputation \\
 & (1) & (2) \\
\midrule
Standard deviation of TWFE estimands & 0.209 & 0.202 \\
 & (0.017) & (0.017) \\
\addlinespace[5pt]
Standard deviation of modern-DiD estimands & 0.231 & 0.232 \\
 & (0.024) & (0.024) \\
\addlinespace[5pt]
Correlation of TWFE and modern-DiD estimands & 0.916 & 0.979 \\
 & (0.023) & (0.004) \\
\addlinespace[5pt]
Prior probability of a sign reversal & 0.115 & 0.054 \\
 & (0.018) & (0.007) \\
\addlinespace[5pt]
Significant wrong-signed TWFE probability & 0.024 & 0.006 \\
 & (0.008) & (0.001) \\
\midrule
Observations & 988 & 1,019 \\
Papers & 89 & 93 \\
\bottomrule
\end{tabular}
\exhibitnotes{This table reports estimates from the Gaussian hierarchical model described in the text, refitted after excluding all pre-treatment comparisons. Estimates come from averaging priors over observed standard errors, giving equal weight to each retained comparison point. Standard deviations are in  regression-sample outcome standard deviations. A sign reversal is $\theta_T\theta_M<0$. A significant wrong-signed TWFE estimate is $\widehat\theta_T\theta_M<0$ and $|\widehat\theta_T|>1.96s_T$, where $\widehat\theta_T\mid\theta_T,s_T\sim N(\theta_T,s_T^2)$. Parentheses report delta-method standard errors clustered by paper, incorporating uncertainty in the fitted parameters and empirical precision mixture. The sample is the same as in Figure~\ref{fig:twfe-baseline-no-pre}. }
\end{table}
\clearpage

\singlespacing
\clearpage
\section{Replication Protocol}\label{app:screening}

\paragraph{Search frame and review.}
The 16,094-record search frame used for this analysis covers the following ten journals and publication windows: the Quarterly Journal of Economics from 2000; the American Economic Review and Econometrica from 2004; the Journal of Political Economy from 2005; the Review of Economic Studies from 2006; the Review of Economics and Statistics, the Journal of the European Economic Association, and the American Economic Journals in Applied Economics and Economic Policy from 2009; and the Economic Journal from 2012. AEJ: Microeconomics and AER: Insights are excluded from this record count and do not contribute papers to the replication sample. Starting years were determined by official journal policy on the posting of replication packages except for the QJE, for which no policy record was found. Papers and Proceedings articles from the American Economic Review were excluded. All search windows ended on October 10, 2026, inclusive; forthcoming papers required evidence that both the accepted, intended-publication manuscript and its corresponding replication package were available by this date. 

Bibliographic records were initially screened using abstracts, keywords, and any publicly available main text, appendices, and replication code. For papers with promising-sounding abstracts but no publicly available drafts, the drafts were manually downloaded with the author’s credentials. Screens use frozen, reproducibly randomized cross-journal queues. Keywords such as ``difference-in-differences,'' ``event study,'' and ``staggered adoption,'' and corresponding software commands were used. Screened-in papers were subjected to further review of main-text tables, figures, equations, and numerical prose and potentially qualifying results were linked to relevant code and data. Most of this review was automated; human oversight and spot-checking were used at various points to ensure fidelity.

\paragraph{Unit of analysis.}
The collection rule was to retain every qualifying main-text specification, including ancillary and diagnostic results and regardless of sign, magnitude, significance, or the authors' causal interpretation. Different outcomes, outcome transformations, samples, treatment definitions, event windows or bins, weights, fixed effects, or controls define distinct specifications. A jointly estimated event-study path counts as one specification.

\paragraph{Eligible designs.}
Both panels and repeated cross sections qualified, provided treatment is assigned to stable units or groups. The typical eligible specification is either a linear regression with unit/group and calendar-time fixed effects and a single binary treatment or an unstacked event study. Time-invariant characteristics interacted with calendar-time fixed effects were allowed, including fixed subgroup-by-time fixed effects. Other time-varying controls, lagged outcomes, unit/group-specific trends, additional treatments, treatment-intensity regressors, subgroup-treatment interactions, and other nonredundant additive fixed effects were excluded. Explicitly identified \citet{callawaysantanna2021} and \citet*{borusyakjaravelspiess2024} estimators provided additional specifications under the same design restrictions. Interacted event studies, as in  \citeauthor{sunabraham2021}, were also permitted when they coincided with \citet{callawaysantanna2021} estimation. Stacked DiD specifications, instrumental variables, nonlinear models, and other estimators were excluded.

\paragraph{Treatment timing and support.}
Adoption dates were established from the original treatment-assignment history. A specification needed to contain at least two distinct first-adoption dates among cohorts with observed untreated and treated outcome observations and an untreated comparison unit in at least one treated period. For a small share of treatments that switch off, we drop periods after adoption in which the original treatment equals zero. The eligibility of a TWFE specification did not require every treated cell to be usable by every modern estimator; estimator-specific support was allowed and incorporated into the leverage screen (detailed below).

\paragraph{Documentation and author outreach.}
Each retained record identified the draft version and exact page, table or figure, panel, row, and column; the outcome, treatment, sample restrictions, weights, fixed effects, and clustering; and the observation, assignment, and calendar-time units. Event studies additionally record bins, omitted periods, and aggregation conventions. These records were randomly spot-checked by the author. 

In the minority of cases where exact replication failed or ambiguities remained about a specification's eligibility, study authors were contacted over email. In some of these cases additional private data or corrected code was shared. A few inexact replications, where the estimated coefficients were qualitatively similar to published figures, were admitted. See below for an accounting of implementation exceptions. 

\paragraph{Leverage screen.}

Each TWFE or modern DiD estimate, constructed as described in the main text, was evaluated for high leverage. To characterize this screen, note first that any such estimate can be written as $\hat\theta=\sum_j a_jY_j$ for some $a_j$. For a clustering partition $c$, define
\begin{equation*}
L_c=\frac{\sum_{j\in c}a_j^2}{\sum_j a_j^2}
\end{equation*}
with $G_{\mathrm{eff}}=(\sum_cL_c^2)^{-1}$, and $L_{\max}=\max_cL_c$. A TWFE and modern DiD pair was retained (in, e.g., Figure \ref{fig:twfe-baseline}) if both estimators have $G_{\mathrm{eff}}\geq10$ and $L_{\max}\leq0.25$. For two-way clustering, these calculations are made in each clustering dimension.

\subsection*{Implementation Exceptions}

\paragraph{Source bin boundaries and a treated reference {\protect\citeprep{bailey2012}}.}

The replication code's event-year bins are retained where the printed text describes different boundaries: annual tails at years $\leq-8$ and $\geq16$, and Table 3 post-treatment bins 1--5, 6--10, 11--14, and $\geq15$. Event year 0 is the omitted treated reference.

\paragraph{Fixed propensity-score donor weights and source county counts {\protect\citeprep{cunninghamgoodmanbacon2025}}.} Estimated propensity scores are treated as fixed weights. Four Table 6 scalar summaries use post-treatment event-window comparisons over years 0--5. 

\paragraph{Exact absorbed nuisance columns {\protect\citeprep{daviswolfram2012}}.}

The age regressor equals a calendar-time component minus a fixed unit-specific operation date, so it is absorbed by unit and calendar-time fixed effects and is dropped.

\paragraph{Bartlett spatial--temporal covariance {\protect\citeprep{deryugina2017}}.}
The source code uses different weights for the spatial-temporal clustering of standard errors than documented in the paper, and results in a variance-covariance matrix that is not positive semidefinite. All 28 specifications use the documented weighting formula.

\paragraph{Fixed-effect specification {\protect\citeprep{giorcellimoser2020}}.}
The source code omits an intercept and one indicator from each fixed-effect set, imposing an additional restriction at the omitted-category intersection. The TWFE comparator uses the full unrestricted additive fixed-effect span.

\paragraph{Recovered OLS first stages {\protect\citeprep{benzellcooke2021}}.}

The two retained specifications are ordinary least squares (OLS) first stages recovered from the original instrumental variable estimation procedure, using its original estimation samples.

\paragraph{Event windows and mixed references {\protect\citeprep*{cantonietal2024,munoz2024}}.}
{\citetrep{munoz2024}} has an adjusted event-study comparison whose post-treatment average uses years 0--6 in the baseline sample and 0--2 in the aligned sample; its separate pretreatment average uses years $-3$ and $-2$. {\citetrep*{cantonietal2024}} retain the published reference bin $(-10,0]$, which includes the adoption year, and subtract its estimated treatment effect in modern contrasts and each bootstrap draw. Unsupported horizons remain unavailable rather than being replaced with zero. 

\paragraph{Pretreatment normalization {\protect\citeprep{lehmannhasemeyerstreb2016}}.}
All units are eventually treated, requiring an additional normalization. The source implementation omits the IPO year and an additional calendar-year indicator; I instead omit event times $-5$ and $-1$, retaining the full firm and year fixed effects and estimating the IPO-year coefficient as a nuisance term. This places both event-time normalizations before treatment. 

\paragraph{Seasonal calendar controls {\protect\citeprep{brehmsaavedra2025,kremeretal2011}}.}
Month-of-year indicators are permitted as additional calendar controls. These are retained alongside state and year fixed effects in the five newspaper-discourse specifications of \cite{brehmsaavedra2025}, and alongside child and survey-round fixed effects in Table IV, column 3 of \cite{kremeretal2011}. 

\subsection*{Imputation Event Studies}
For imputation estimation of standard event studies, I follow \cite{borusyakjaravelspiess2024} in regressing outcomes on the allowed fixed effects, controls, and lead-bin indicators using untreated observations, then average the identified lead coefficients over the bins. In settings with an explicit reference period, I instead average $Y_j-\hat Y_j(0)$ within each supported bin, average the bins equally, and subtract the weighted residual mean in the source reference observations. This includes wholly untreated reference sets. The latter pretreatment summaries are in-sample residual contrasts, not standard \cite{borusyakjaravelspiess2024} pretrend coefficients; see \cite{roth2026eventstudies} for discussion of these issues. 

\begingroup
\singlespacing
\setlength{\bibsep}{5pt plus 1pt}
\interlinepenalty=10000
\nociterep{abebereseetal2023,acemogluazar2020,ageretal2020,aghionetal2008,agrawalgoldfarb2008,akerlofetal2026,
  amanranaminaudier2026,andrews2023,avisetal2018,bahajreis2026,bailey2012,baileygoodmanbacon2015,
  bancalari2026,banerjeeetal2020,baumsnowlutz2011,bazzietal2026,beckeretal2025,bellegodrouard2024,
  benzellcooke2021,bertrandetal2021,bhalotraetal2023,borcanetal2017,bragamalkova2024,braguinskyetal2015,
  buchmuellercarey2018,calamuncilonsky2025,cantonietal2017,cantonietal2024,cantonipons2021,cervellatietal2017,
  chenetal2022,chiovellietal2025,chuetal2021,cicala2015,cohenetal2011,cohenetal2017,
  colonnelliprem2022,conleytaber2011,cookshah2022,coubeetal2026,couttenieretal2017,dahisszerman2026,
  dalboetal2009,daviswolfram2012,dejanvryetal2015,deryugina2017,dippelheblich2021,donaldson2018,
  egedesoetal2020,espositoetal2023,falcone2026,fasanietal2021,fetzer2022,fetzeretal2021,
  frakes2013,funkeetal2023,galianietal2005,gibsonmckenzie2014,giorcellimoser2020,gokmenetal2021,
  goncalvesetal2026,gonzalezetal2024,hamoryetal2021,hansmanetal2020,hewang2017,hjortpoulsen2019,
  hodlerraschky2014,hollingsworthetal2024,janssenzhang2023,jensenmiller2018,jiaetal2015,johnsonetal2024,
  johnsonnoguera2017,kreindleretal2026,laferraraetal2012,lewisfaupeletal2016,martinezbravo2017,martinezbravoetal2022,
  mastroroccoteso2026,moseretal2014,munoz2024,noackcostello2024,norrisxiong2023,priks2015,
  rexer2025,sanchezdelasierraetal2024,shin2025,snyderwelch2017,spenkuchetal2023,stevensonwolfers2006,
  sunyannelis2016,weill2026,wolfers2006,zussman2023,
  bindlerhjalmarsson2018,chenlan2017,iyeretal2012,lebarbanchonsauvagnat2022,anderssonetal2023,
  bostwicketal2022,carrilloetal2023,nagengastyotov2025,yangetal2024,gibson2019,
  lafortuneetal2018,cunninghamgoodmanbacon2025,boomhower2019,yousaf2021,askarovetal2023,
  lehmannhasemeyerstreb2016,beg2022,brehmsaavedra2025,chirakijja2024,kremeretal2011,
  ashgallettagiommoni2025,
  starcwollmann2025}
\bibliographystylerep{aer}
\renewcommand{\bibsection}{\section*{\refname}}
\bibliographyrep{indifference_refs}
\endgroup

\end{document}